\documentclass[12pt]{JHEP3}

\def\nn{\nonumber}
\def\be{\begin{eqnarray}}
\def\ee{\end{eqnarray}}

\def\l{\left}
\def\r{\right}
\def\hp{\hat{+}}
\def\hm{\hat{-}}
\def\hi{\hat{i}}
\def\hj{\hat{j}}
\def\hon{\hat{1}}
\def\htw{\hat{2}}
\def\hth{\hat{3}}
\def\hfo{\hat{4}}
\def\hfi{\hat{5}}

\def\123{\hat{1}\hat{2}\hat{3}}
\def\49{\hat{4}\hat{5}\hat{6}\hat{7}\hat{8}\hat{9}}

\def\xxx#1 {{\sf hep-th/#1} }
\def\IR{\relax{\rm I\kern-.18em R}}

\title{Supersymmetry and Branes in M-theory Plane-waves}

\author{Nakwoo Kim \\
Max-Planck-Institut f\"ur Gravitationsphysik\\
Albert-Einstein-Institut\\
Am M\"uhlenberg 1, D-14476 Golm, Germany\\
{\tt kim@aei.mpg.de}
}

\author{Jung-Tay Yee  \\
School of Physics, Korea Institute for Advanced Study\\
207-43, Cheongryangri-Dong, Dongdaemun-Gu, Seoul 130-012, Korea\\
{\tt jungtay@kias.re.kr} }

\preprint{hep-th/0211029 \\ AEI-2002-088 \\ KIAS-P02067}
\abstract{We study brane embeddings in M-theory plane-waves 
and their supersymmetry. The relation with
branes in AdS backgrounds via the Penrose limit
is also explored. Longitudinal planar branes are originated from AdS branes
while giant gravitons of AdS spaces become spherical branes which 
are realized as fuzzy spheres in the massive matrix theory.}

\begin{document}

\section{Introduction}
The physics of IIB string theory and M-theory in the maximally
supersymmetric plane wave backgrounds \cite{ppwave} turns out to be 
surprisingly rich. 
In the light-cone gauge the superstring and the supermembrane
Green-Schwarz actions both significantly simplify. 
The string worldsheet theory has free massive bosons
and fermions, and the free string light cone spectrum is known exactly
\cite{metsaev}. The supermembrane action is already interacting
in the flat background, and the gravitational wave adds two new
types of terms to the light-cone action: mass terms and bosonic
cubic interaction terms \cite{bmn}. 
It is well known that the light-cone supermembrane action 
can be discretized to give the Yang-Mills quantum
mechanics \cite{dehoni}, which is usually called ``Matrix theory'' 
providing a non-perturbative partonic description of M-theory \cite{bfss}. 
In relation with IIA string theory the cubic interaction terms
are easily identified as describing the Myers' dielectric effect 
\cite{myers}: the constituent D0-branes are expanding into fuzzy spheres. 
Let us quote here the plane-wave solution of eleven dimensional
supergravity which is of utmost interest in this paper,
\be
 ds^2 &=& -4 dx^+ dx^- - \l[\l( {\mu \over 3} \r)^2 y^2 + \l({\mu
  \over 6} \r)^2 z^2 \r] dx^{+2} + d \vec{y}^2 + d\vec{z}^2 \\ \nn
 F &=& \mu \, dx^+ \wedge dy^1 \wedge dy^2\wedge dy^3 ,
\ee
where $\vec{y},\vec{z}$ are vectors in $\IR^3,\IR^6$ respectively.
The matrix theory in this particular background is first given in \cite{bmn},
and the derivation by discretizing the supermembrane action is 
demonstrated in \cite{dasraa1}.
One notable feature of this solution is that already at the
level of metric the symmetry of the nine dimensional transverse
space is broken to $SO(3)\times SO(6)$.
The existence of a dimensionful parameter $\mu$ renders the
study of matrix model in some sense even more tractable than
the flat space counterpart. In the original matrix theory 
a perturbative approach is hard to achieve first because of
continuous moduli and secondly due to the lack of dimensionless parameter.
Now with the plane-wave matrix theory the moduli space is a discrete
set of fuzzy spheres of different radii, and there exists a 
dimensionless coupling constant which makes perturbative calculations
possible \cite{dasraa1}. 
By exploiting the fact that the symmetry algebra contains
a classical Lie superalgebra $SU(2|4)$ and studying its atypical, i.e.
short representations, it is shown that there exist
protected states whose energies are free from perturbative corrections
\cite{dasraa2, kimple, kimpark}.

The aim of this letter is to provide a list of supersymmetric
branes in the eleven dimensional plane-waves through supergravity analysis. 
It can be taken as the M-theory answer to the
paper by Skenderis and Taylor \cite{sketay} who studied supersymmetric
D-branes in $AdS_5\times S^5$ and the plane-wave backgrounds of IIB
string theory. The motivation for such a study is obvious when we
recall the importance of D-branes in modern string theory. Especially
in terms of the AdS/CFT correspondence \cite{adscft}, the branes correspond to
several interesting objects like magnetic monopoles, baryonic vertex
\cite{baryon},
giant gravitons \cite{ggrav} and defect conformal field theory 
\cite{dcft}. The supergravity analysis of \cite{sketay, chuho} is found to 
agree with microscopic constructions of
D-branes as open string boundary conditions \cite{dabho} and
as squeezed states of closed string sector \cite{billo}. 
These 1/2-BPS branes are also constructed as localized supergravity
solutions in \cite{baimee}.
For M-theory a comparison can be made with 
the matrix model constructed in \cite{bmn}, where 
1/2-BPS fuzzy sphere solutions are presented. A systematic
search of supersymmetric branes as matrix theory solitons
is undertaken in \cite{bak, jhpark}, and a new matrix model of
fivebranes in plane-wave is constructed in \cite{leeyi}
as ${\cal N}=8$ gauge quantum mechanics with extra hypermultiplets.  
We find our result consistent with the literature as it should be.
For related works on eleven dimensional plane-wave solutions
see \cite{others}.

The particular form of plane-wave
makes it natural to classify branes first according to the
behaviour in terms of $x^+$ and $x^-$. We will be interested
in the branes which are extended along both $x^+$ and $x^-$, and 
also the branes which are extended along $x^+$ while localized 
in $x^-$. We will call them ``longitudinal'' and ``transverse''
branes respectively. In the matrix theory description the longitudinal
M5-branes are realized as four dimensional objects, while it is the
transverse spherical M2-branes which become fuzzy spheres of matrix theory. 
For completeness we will also present longitudinal M2-branes and
transverse M5-branes in the plane wave as well, although they are
not immediately related to the solitons of matrix theory.

It is by now well established that the plane-waves are the Penrose
limits \cite{penrose} of AdS solutions. In the Penrose limit the spacetime 
is blown up around the worldline of a chosen null geodesic. For
the case of AdS backgrounds, if the massless particle moves
in the sphere the limits are 
plane-waves, while for particles moving only in AdS the spacetime becomes
Minkowski. Now an interesting question is what happens to the
supersymmetric branes in AdS space in the Penrose limit. 
There are two types of half supersymmetric branes in AdS
backgrounds: AdS branes and giant gravitons. In order to
get AdS branes it is convenient to start with intersecting
branes configurations and take the near horizon limit of one
brane. Let us take M5 and M2 branes intersecting on a string 
as an example. This system preserves 8 supersymmetries. When
we take the near horizon limit of M2-branes the background
geometry becomes $AdS_4\times S^7$, and likewise the 6-dimensional 
M5-brane worldvolume occupies $AdS_3$ subspace of $AdS_4$ and
$S^3$ inside $S^7$. The supersymmetry is enhanced to 16,
and one can conjecture that this configuration is dual to a two dimensional 
superconformal field theory with $SO(4)$ global symmetry. 
Similar configurations are summarised in the Table 1.
\begin{table}
\begin{center}
\begin{tabular}{|c|c|c|c|}
\hline
Brane & Intersection & AdS embedding & pp-wave embedding
\\ \hline
M2 & $(0| M2 \perp M2)$ & $AdS_2\times S_1$ & $(+,-,1,0)$
\\ \hline
M2 & $(1| M2 \perp M5)$ & $AdS_3$ & -
\\ \hline
M5 & $(1| M5 \perp M2)$ & $AdS_3\times S_3$ & $(+,-,2,2)$
\\ \hline
M5 & $(3| M5 \perp M5)$ & $AdS_5\times S_1$ & $(+,-,0,4)$
\\ \hline
M5 & $(1| M5 \perp M5)$ & $AdS_3\times S_3$ & $(+,-,2,2)$
\\ \hline
\end{tabular}
\caption{AdS branes and the corresponding planar branes in the 
plane-wave.}
\end{center}
\vspace{-0.5cm}
\end{table}
The giant gravitons are spherical M2 or M5 orbiting at light velocity 
in $S^4$ or $S^7$.
If we choose null geodesics moving along with the giant gravitons
the brane geometry is kept intact. They become transverse spherical branes
in plane-waves.

In the main part of this paper we use $\kappa$ symmetric
membrane and fivebrane actions to check the equation of
motion and supersymmetry of various brane embeddings in plane-waves. 
We choose to use the same notation which is introduced in \cite{sketay} 
to denote different brane configurations. 
Worldvolume directions are given in the parenthesis, so for instance 
$(+,-,2,2)$ means fivebranes 
extended along $x^+,x^-$, and two directions in $\IR^3$ and $\IR^6$ each.
When one checks the supersymmetry it is essential to have the explicit
form of Killing spinors. They can be found in the literature but
for completeness and to set the notation we show the derivations 
in the appendix.

\section{M2-branes in plane-waves}
It is a relatively simple matter to check the supersymmetry of membranes
with simple geometries in the plane wave background. 
The bosonic part of the supermembrane action can be written as
\cite{bersez}, 
\be
 S= - T \int d^3 \sigma 
\l( \sqrt{-\det g} - C  \r),
\ee
where $g$ and $C$ are the induced metric and 
the three-form gauge field pulled back on the worldvolume respectively.
Unbroken supersymmetry requires that the Killing spinors of 
the background geometry be consistent with the so-called
$\kappa$-symmetry projections, so
\be
\Gamma_\kappa \epsilon = \epsilon , 
\ee
where
\be
\Gamma_\kappa = \frac{1}{3!} \epsilon^{mnp}
\partial_m X^M
\partial_n X^N
\partial_p X^P
\Gamma_{MNP} .
\ee
Capital latin letters denote eleven dimensional indices and
lowercase is reserved for worldvolume indices. The equation of 
motion is written as follows,
\be
 {1 \over \sqrt{-g} } 
  \partial_m \l( \sqrt{-g} g^{mn} \partial_n X^N \r)G_{MN} 
  +g^{mn} \partial_m X^N \partial_n X^P \gamma_{MNP}
  = 
  {1 \over 3!} \epsilon^{mnp} F_{Mmnp}, 
\ee 
where $G$ is background metric, $\gamma_{MNP}$ are Christoffel 
symbols and $F= dC$. 

\subsection{Longitudinal branes}
These branes are one dimensional in the transverse nine dimensional
space, and it is straightforward to see that they satisfy the
equation of motion with linear geometries. When lying along the $i$-th
direction (hatted indices represent the tangent space), 
\be
 \Gamma_\kappa = \Gamma_{\hp\hm\hat{i}}.
\ee
and in order for the projection to be satisfied at every point 
of $x^+$ the conditions
\be
 &&\Gamma_{\hp\hm\hat{i}} \epsilon_0 = \epsilon_0 \\
 && \Gamma_{\hp\hm\hat{i}} \Gamma_{\hon\htw\hth} \epsilon_0
    = \epsilon_0
\ee
have to be fulfilled.
We find that if the membrane is extended purely in $\IR^3$ it has at least 
1/4-supersymmetry and the supersymmetry is enhanced to one half 
when located at the origin. And the longitudinal membranes lying
in $\IR^6$ direction break the supersymmetry completely. It is
essentially because $\Gamma_\kappa$ anti-commutes with
$\Gamma_{\hon\htw\hth}$.
\subsection{Transverse branes}
Another class of supersymmetric membranes have spherical geometry
in $\IR^3$. The equations of motion for transverse scalars $z^{a'}$ 
force them to lie at the origin, and the radius $r$ of the sphere is 
found to be arbitrary. We further get 
\be
 \Gamma_\kappa = {3 \over {\mu r}} 
 \l( \Gamma_{\hp} 
 -{r^2\over 2} 
  \l( {\mu \over 3}\r)^2 
  \Gamma_{\hm} \r) 
  \Gamma_{\123} \Gamma_{\hat{a}} y^a.
\ee
It turns out that $\Gamma_\kappa \epsilon = \epsilon$ is satisfied 
for any $x^+$ and $r$, for $\Gamma_{\hp} \epsilon_0 =0.$ 
This is precisely the projection of Killing spinors which are 
linearly realized in the supermembrane action in the light-cone
gauge or the matrix quantum mechanics. And it agrees with the observation
in the matrix theory that the fuzzy sphere solutions in $\IR^3$
preserve the whole linearly realized supersymmetry while breaking
the nonlinearly realized supersymmetries completely.

There also exist transverse branes of planar geometry. The
equations of motion are satisfied for $(+,1,1)$ and $(+,0,2)$
branes. Due to Wess-Zumino couplings $(+,2,0)$ planar branes do not satisfy
the equation of motion without transverse scalar excitations. None
of these planar transverse branes are supersymmetric.
\section{M5-branes in plane-waves}
\subsection{Introduction to PST formulation of fivebrane action}
M-theory fivebranes and the gauge field theory confined on 
their worldvolume are certainly one of the most mysterious objects
in string theory. The construction of covariant action is a
subtlety because of the selfdual three-form field strength.
There exist several proposals for M5 brane actions in the 
literature \cite{pst,howsez,schwarz}. Among them, 
covariant field equations from superembedding approach 
\cite{howsez} is proven to be equivalent to other approaches
\cite{banlec}. Noncovaraint action of \cite{schwarz} 
can be obtained from \cite{pst} with gauge fixing of auxiliary field. 
In this paper we use Pasti, Sorokin and Tonin (PST) \cite{pst,soroki} 
formulation which is manifestly covariant. In this section, we review 
\cite{soroki} briefly.  

The bosonic part of PST action is 
\be
 S &=& T_{M5} \int_{{\cal M}_6} 
     d^6 x \l[ - \sqrt{-\det \l( g_{mn} + \tilde{H}_{mn}\r) } +
  {1\over 4} \sqrt{-g} \tilde{H}^{mn} H_{mn}  \r.
\nonumber \\
 && \l. - {1\over 2} \l( C_6 + dA_2 \wedge C_3  \r) \r],
 \label{pstaction}
\ee
where $g_{mn}$ is the induced metric on the worldvolume. $C_6$ and $C_3$ are
pullback of Ramond-Ramond potentials which are subject to 
the eleven dimensional Hodge duality condition : 
\be
 F_7 = dC_6 - C_3 \wedge dC_3 = * F_4.
\ee
$A_2$ is worldvolume gauge field, which gives modified field strength
on the worldvolume,  
\be
 H_3 = dA_2 - C_3.
\ee
There is an auxiliary scalar field $a(x)$ such that 
\be
 H_{mn} = H_{mnp} v^p, \qquad \tilde{H}_{mn}= \tilde{H}_{mnp} v^p, \qquad
 {\rm  with} \quad v_p = {\partial_p a \over \sqrt{-g^{mn} \, \partial_m a \, 
  \partial_n a} },
\ee
where $\tilde{H}_3$ is Hodge dual to $H_3$ on the worldvolume :
\be
 \tilde{H}^{mnp} = {1 \over 3!} \epsilon^{mnpijk} H_{ijk}.
\ee
It can be shown that upon double dimensional reduction, one
obtains dual form of 
D4-brane action and $\tilde{H}_{mn}$ reduces to gauge field strength on the 
worldvolume.  

The PST action has the following four different gauge symmetries
\be
 &&1. \quad \delta A_2 = d \Lambda, \nn \\
 &&2. \quad \delta A_2 = d a \wedge \phi , \qquad \delta a=0, \nn \\
 &&3. \quad \delta a = \varphi, \qquad \delta A_{mn} = {\varphi \over
  \sqrt{-(\partial a)^2}}\l( {2 \delta L_{\rm BI} \over \delta \tilde{H}^{mn} 
  } - H_{mn} \r), \nn\\
 &&4. \quad \delta A_2 = B_2, \qquad \delta C_3 = dB_2.   
\ee
Here $L_{\rm BI} \equiv \sqrt{\det\l(\delta_m^{~n} + \tilde{H}_m^{~n}\r)}$. 
Note that the first symmetry is the same as the usual gauge symmetry of 
Dirac-Born-Infeld action of D-branes and the fourth one is simply a 
pullback of eleven dimensional gauge symmetry.
Upon gauge fixing of scalar field $a$ , for example, as $a = x^5$, 
we obtain the noncovariant formulation of \cite{schwarz}.
From the equation of motion of $A_2$,
self-duality constraint is incorporated automatically : 
\be
\label{nlsd}
 H_{mn} = {\tilde{H}_{mn} - 1/2 \, {\rm tr} \tilde{H}^2 \tilde{H}_{mn}
  +\tilde{H}^3_{mn} \over L_{\rm BI}}. 
\ee
Equation of motion for $X^M$ is 
\be
 &&  \epsilon^{m_1 \cdots m_6 } \l( {1\over 6!} 
    F^{\hat{m}}{}_{m_1 \cdots m_6} 
   -{1 \over (3!)^2} \l(F^{\hat{m}}{}_{m_1 m_2 m_3} H_{m_4 m_5 m_6} - 
  \partial_n X^{\hat{m}} F^n{}_{m_1 m_2 m_3} H_{m_4 m_5 m_6} \r) \r) 
\nn\\
 &&  \quad\quad =-{1 \over 2}  T^{mn} \nabla_m \partial_n X^{\hat{m}} 
\ee
where
\be
 T_{mn} = 2 g^{mn} \l(L_{\rm BI} - {1 \over 4} \tilde{H}_{mn} H^{mn} \r)
   -{1 \over 2} H^{mpq} \tilde{H}^n_{pq}.
\ee
In order to incorporate fermions and make the action supersymmetric
one replaces the fields and coordinates by superforms and supercoordinates.
The $\kappa$ symmetry is more involved than that of membranes 
because of the gauge field and the auxiliary scalar $a$. 
\be
 \Gamma_\kappa &=& - {v_m \Gamma^m \over \sqrt{-\det(g + \tilde{H})} }
   \l( {1\over 5!} \epsilon^{i_1 \cdots i_5 n} \Gamma_{i_1 \cdots i_5} v_n 
   + {1\over 2}\sqrt{-g} \Gamma_{np} \tilde{H}^{np} \r.  \nonumber\\
&& 
\l. \quad\quad\quad
 + {1 \over 8} \epsilon^{m n_1 n_2 p_1 p_2 q} \Gamma_m \tilde{H}_{n_1 n_2}
  \tilde{H}_{p_1 p_2} v_q \r).
\ee
$\Gamma_m = e_m{}^{\hat{m}} \Gamma_{\hat{m}} $ is pullback of eleven 
dimensional gamma matrices on six dimensional worldvolume. 
One can check that $\Gamma_\kappa$ as given above 
is traceless and squares to identity.

\subsection{Longitudinal branes}
\subsubsection{$(+,-,3,1)$ branes}
We notice that there is a source term to the worldvolume
flux from the Wess-Zumino coupling to the background 4-form 
fields. This phenomenon is essentially the same as the M5-brane 
baryonic vertices in $AdS_7\times S^4$ \cite{gomram}, and one could start 
from the configurations in the AdS background and take the Penrose limit, 
but here we will derive the general
solutions of brane equation of motion. We set the notation
for the null fluxes as (we take $z^4$ to be along the worldvolume)
\be
\tilde{H} = \frac{1}{2} dx^+ \wedge dy^a \wedge dy^b f^c \epsilon_{abc}
          + dx^+ \wedge dy^a \wedge dz^4 g_a, 
\ee
then from the Bianchi identity we get
\be
\partial_a f^a &=& \mu
\nonumber\\
\epsilon^{abc} \partial_b g_c &=& 0
\ee
Now when we choose $a=z^4$, it is straightforward to get
\be
H_{+a} &=& - i g_a 
\nonumber\\
\tilde{H}_{+a} &=& - i f_a
\ee
and when they are substituted into the generalized self-duality
equation eq.(\ref{nlsd}) it gives simply
\be
f_a = g_a
\ee
In fact when we evaluate the nonlinear terms in the equation we
find they vanish. This is not unexpected since the fluxes are
null and higher order Lorentz invariants constructed
by contracting indices typically vanish. One simple solution is given 
as $f_a=g_a= \frac{\mu}{3} x^a$. Now we can check whether 
these branes are supersymmetric, which means $\Gamma_\kappa\epsilon
=\epsilon$ should be satisfied everywhere on the worldvolume.
When we spell out the required conditions we find it is impossible
to satisfy especially at every $x^+$. Essentially the
reason is $\Gamma_{\hp\hm\hon\htw\hth\hfo}$ does not commute with
$\Gamma_{\hon\htw\hth}$ which dictates the $x^+$-dependence
of all Killing spinors. Similar objects in IIB plane waves are
$(+,-,4,0)$ supersymmetric D5-branes with null fluxes turned on due
to Wess-Zumino couplings, so the analogy does not persist here. 
We give more comments on this issue in section 4.

\subsubsection{$(+,-,2,2)$ branes}
For this type of branes the pull back of three-form field vanishes
so the worldvolume gauge field $\tilde{H}$ can be set to zero.
For clarity let us choose $y^1,y^2,z^4,z^5$
to be worldvolume directions. Using $\Gamma_{\kappa}= 
\Gamma_{\hp\hm\hon\htw\hfo\hfi}$, the projection condition gives
the following equations
\be
\label{1245}
 &&\Gamma_{\hp\hm\hon\htw\hfo\hfi} Q \epsilon_0 = Q \epsilon_0 \\
 &&\Gamma_{\hp\hm\hon\htw\hfo\hfi} Q \Gamma_{\hp\hm\hon\htw\hth} \epsilon_0
  = Q \Gamma_{\hp\hm\hon\htw\hth} \epsilon_0 \\
 &&\Gamma_{\hp\hm\hon\htw\hfo\hfi} Q \Gamma_{\hon\htw\hth} \epsilon_0 =
   Q \Gamma_{\hon\htw\hth} \epsilon_0 \\
 &&\Gamma_{\hp\hm\hon\htw\hfo\hfi} Q\Gamma_{\hp\hm} \epsilon_0
   = Q\Gamma_{\hp\hm} \epsilon_0,
\ee
where
\be
 Q \equiv \l(1 + {\mu \over 6} y^a \Gamma_{\hat{a}}\Gamma_{\hm\hon\htw\hth}
     - {\mu \over 12} z^{a'} \Gamma_{\hat{a'}} \Gamma_{\hm\hon\htw\hth}
         \r).
\ee
Now using the commutation properties of matrices involved, it is 
straightforward to see that eq.(\ref{1245}) is satisfied if
the transverse scalars are set to zero 
and
\be
\label{22proj}
 \Gamma_{\hp\hm\hon\htw\hfo\hfi} \epsilon_0 = \epsilon_0,
\ee
implying that $(+,-,2,2)$ branes are 1/2-BPS when they sit at the
origin. For the branes located away from the origin, they still
preserve 1/4 of the supersymmetries for the Killing spinors
which are annihilated by $\Gamma_{\hat{-}}$.

We can also consider nonzero gauge fields on the worldvolume. 
We will see that turning on null fluxes $H_{+45}=H_{+12}$ does
not break the supersymmetry provided the brane is accordingly 
moved away from the origin. The origin of such worldvolume fields
in AdS backgrounds is not hard to find. $(+,-,2,2)$ branes
are Penrose limits of $AdS_3\times S^3$ branes, where nonzero three-form
flux can be turned on through $AdS_3$ and $S^3$. When the Penrose
limit is taken the flux becomes null just like the background four-form
field. This example is similar to $AdS_4\times S^2$ 
D5-branes with nonzero flux through $S^2$ which was explicitly
studied in \cite{sketay}.

We have 
$\Gamma_{\kappa}= \Gamma_{\hp\hm\hon\htw\hfo\hfi} - H_{+45}
  \Gamma_{\hm\hon\htw}$, 
and the branes are supersymmetric if
\be
\label{1245gauge}
 &&\Gamma_\kappa Q \epsilon_0 = Q \epsilon_0 \\
 &&\Gamma_\kappa Q \Gamma_{\hp\hm\hon\htw\hth} \epsilon_0 = Q
  \Gamma_{\hp\hm\hon\htw\hth} \epsilon_0 \\
 &&\Gamma_\kappa Q \Gamma_{\hp\hm\hon\htw\hth} \epsilon_0 = Q
   \Gamma_{\hon\htw\hth} \epsilon_0
\ee
are satisfied. There are at least 1/4 supersymmetries 
with $\Gamma_{\hm}\epsilon_0 = 0$, and the supersymmetry
is enhanced to 1/2, with eq.(\ref{22proj}) as the projection rule 
and 
\be
 && H_{+45} = {\mu \over 3} y^3 ,
\ee
with other transverse scalars set to zero.

\subsubsection{$(+,-,1,3)$ branes}
For this orientation the branes are not supersymmetric
irrespective of positions.

\subsubsection{$(+,-,0,4)$ branes}
The analysis is similar to that of $(+,-,2,2)$ branes. 
The branes are 1/2-BPS at the origin and 1/4-BPS away from it. 
One might ask whether it is also possible to turn on worldvolume
flux and move the branes away from the origin, like $(+,-,2,2)$ branes. 
When one proceeds for instance with nonzero $H_{+67}=H_{+45}$ 
one finds that there is no relation between the flux and
the position of the brane like $(+,-,2,2)$ branes.
So it is not possible to compensate the harmonic potential with
null worldvolume fluxes in this case.

\subsection{Transverse branes}
The consideration is analogous to the spherical M2-branes in 
$\IR^3$. The effective harmonic potential of light-cone gauge
action puts the five dimensional sphere of arbitrary radius
at the origin of $\IR^3$, and we have
\be
 \Gamma_\kappa = {6 \over {\mu r}} 
 \l( \Gamma_{\hp} 
 -{r^2\over 2} 
  \l( {\mu \over 6}\r)^2 
  \Gamma_{\hm} \r) 
  \Gamma_{\49} \Gamma_{\hat{a'}} z^{a'}.
\ee
The projection condition is again satisfied provided 
$\Gamma_{\hat{+}}\epsilon_0=0$. They can be traced back
to $AdS_4\times S^7$ backgrounds in the same way: as giant gravitons
or M5-branes orbiting $S^7$. Unlike spherical membranes,
these solutions are not realized as solitons of 
the massive matrix model in \cite{bmn}. This is not unrelated to 
the well-known difficulty of constructing odd dimensional
objects in matrix models.

The study of transverse planar M5-branes is again similar to that
of transverse planar M2-branes. Due to Wess-Zumino couplings
$(+,3,2)$ should have gauge fields while transverse scalar
field has to be turned on in $(+,0,5)$ branes. $(+,2,3)$
and $(+,1,4)$ branes satisfy the equations of motion without
field excitations. They are all non-supersymmetric.

\section{Discussions}
In this paper we have employed $\kappa$ symmetric membrane and fivebrane 
actions to find supersymmetric branes in eleven dimensional plane waves.
The result is consistent with the predictions based on known
supersymmetric brane configurations in AdS backgrounds,
and the next step is naturally to compare with the branes found
in the matrix theory. From the matrix equation of motion one 
readily sees that the 
mass terms invalidate the planar membrane solutions of ordinary
matrix theory in flat space, let alone supersymmetry. 
Membranes with rather nontrivial
geometries such as hyperbolic surfaces can be found instead \cite{bak}.
The non-supersymmetric transverse planar membranes reported in this
paper should not be taken as contradictory with matrix theory
results. The matrix theory is obtained after light-cone gauge
fixing, and $x^-$ is not at our disposal but determined by the
Virasoro constraints. In this work $x^-$ is always set to a constant
for transverse branes. Usually the constraint equation does not allow us to
set $x^-$ to a constant, but for transverse spherical membranes the
constraint equation becomes trivial and that is why two approaches
coincide. 

An alternative to soliton description is possible with fivebranes
in matrix theory. The open string modes between D0 and D4-branes in IIA
string picture \cite{berdou} give rise to hypermultiplets in the
matrix quantum mechanics. The plane wave deformation of this matrix theory
is presented in \cite{leeyi} for $(+,-,2,2)$ branes in our notation,
and certainly it will be interesting to construct the matrix theory
of $(+,-,0,4)$ branes which are also supersymmetric.

By and large our result goes hand in hand with IIB
branes in $AdS_5\times S^5$ and plane-waves. Especially with
AdS branes and giant gravitons we find perfect analogy, so we
look for other pairs of supersymmetric branes 
in ten and eleven dimensional plane waves. We are especially
interested in two types of IIB branes in plane waves which 
have the peculiarity that supersymmetries do not depend where
they are located. Curiously we have not found similar objects in M-theory
plane waves. 

Firstly there exist D-strings from unstable D-strings in $AdS_5\times S^5$ 
which are wrapped on a great circle of $S^5$ \cite{baipee}. 
The supersymmetry is enhanced under Penrose limit and these
D-strings have 8 supersymmetries in plane waves everywhere in $\IR^8$.
For these 1/4-BPS D-strings, the analogy might be
membranes wrapping $S^2$ of $S^4$ in $AdS_7\times S^4$.
This configuration satisfies the equation of motion, but surely this
is not supersymmetric\footnote{We can check explicitly using Killing spinors 
in global coordinates presented in appendix}. 
If we take the Penrose limit, the result should 
be $(+,-,1,0)$ membranes with enhanced symmetries. Just up to this point,
situations seem to be the similar to type IIB case, but this type of
membranes are 1/2-BPS at the origin differently from D-strings.
And in fact $(+,-,1,0)$ can be obtained from $AdS_2\times S_1$
membranes as presented in table 1.

We are also interested in $(+,-,4,0)$ D5-branes with null worldvolume 
flux turned on by Wess-Zumino couplings. Once the gauge field is 
turned on they have 16 supersymmetries irrespective of positions. 
It is $(+,-,3,1)$ M5-branes which can be matched with $(+,-,4,0)$
D5-branes but according to our analysis these M5-branes are not
supersymmetric. In our opinion this does not contradict the known baryonic
M5-branes wrapping $S^4$ in $AdS_7\times S^4$ which are supersymmetric
\cite{gomram}. 
The Penrose limits of baryonic D5-branes are studied in 
\cite{ppbary}, where it is illustrated that resulting configuration 
is localized in $x^+$
which originates from the affine parameter of the null geodesic.
It is because the null geodesics have a nonvanishing component along the
radial direction of Poincare coordinate system, so they intersect
with the brane worldvolume at a point. 
Static configurations in global coordinates
could give longitudinal branes because massless particles moving purely
in $S^5$ can be chosen \footnote{On further details of choosing
null geodesics when one takes the Penrose
limits in different coordinate systems we refer the readers to
\cite{blau}.}, but the baryonic branes in the literature are all constructed
in Poincare coordinates. Unfortunately finding supersymmetric baryonic 
branes which are static in global coordinates does not seem promising. 
In global coordinates the Killing spinors depend on all coordinates 
\footnote{We could not find the explicit form of Killing
spinors in global coordinates from the literature, so decided to present
the solutions in appendix.}, so satisfying $\Gamma_\kappa$ 
projection everywhere on the worldvolume is more difficult.  
In fact it is not even clear how to put background D3, M2 or M5-branes 
supersymmetrically.

We thus conclude that the analogy between IIB and M-theory is 
not extended beyond AdS branes and giant gravitons. It might
simply mean that the Penrose limit acts differently with
different AdS solutions, but one cannot rule out the possibility
that D-strings and $(+,-,4,0)$ D5-branes are in fact spurious 
and unphysical. We think it is an important matter to
check their consistency for instance following the approach 
advocated in \cite{bergab}.

{\bf Note added:} After this paper was completed we received an
interesting paper by Skenderis and Taylor \cite{sketay2}, 
where open string boundary conditions for light-cone worldsheet 
action in IIB plane-waves is carefully re-investigated. 
It is argued that one can restore some of the broken spacetime 
supersymmetries by using worldsheet symmetries. 
It will be very exciting to check whether such additional symmetries 
can be found also for branes in M-theory plane waves.

\section*{Acknowledgments}
We would like to thank Hwang-hyun Kwon, Kimyeong Lee, Yoji Michishita, 
Jeong-Hyuck Park, Soo-Jong Rey and Piljin Yi for useful discussions. 
The work of NK is supported by German Research Foundation (DFG).
NK would like to thank KIAS for hospitality where part of this work
was done.
\begin{appendix}
\section{The derivation of Killing spinors}
\subsection{Plane-waves}
An explicit derivation of the Killing spinors in eleven dimensional
plane wave can be also found in \cite{josgeo}.
The pp-wave solution of interest in this paper is given as 
\be
 ds^2 &=& -4 dx^+ dx^- - \l[\l( {\mu \over 3} \r)^2 y^2 + \l({\mu
  \over 6} \r)^2 z^2 \r] dx^{+2} + d \vec{y}^2 + d\vec{z}^2 \\ \nn
 F_{+123}&=& \mu,
\ee
where indices of $y^a$ and $z^{a'}$ are $a=1,2,3$ and $a'=4,5,6,7,8,9$ for
each. We define the tangent space as follows,
\be
 &&\eta_{\hp \hm} = \eta_{\hp\hm} = 1, \quad \eta_{\hp\hm}=\eta_{\hm\hm}=0,
  \quad \eta_{\hi\hj}= \delta_{ij} \\ \nn
 && e_-{}^{\hm} =2, e_+{}^{\hm} = {1\over 2}
 \l[\l( {\mu \over 3} \r)^2 y^2 + \l({\mu \over 6} \r)^2 z^2 \r] ,
 e_+{}^{\hp} = -1, e_i{}^{\hj} = \delta_i{}^j
\ee

Thus,
\be
 \Gamma^- &=& {1 \over 2} \Gamma_{\hp} + {1\over 4} \l[\l( {\mu \over 3} \r)^2 y^2 + \l({\mu
   \over 6} \r)^2 z^2 \r] \Gamma_{\hm }\\ \nn
 \Gamma^+ &=& - \Gamma_{\hm} \\ \nn
 \Gamma_- &=& 2 \Gamma_{\hm} \\ \nn
 \Gamma_+ &=& - \Gamma_{\hp} + {1 \over 2} \l[\l( {\mu \over 3} \r)^2 y^2 + \l({\mu
   \over 6} \r)^2 z^2 \r] \Gamma_{\hm}
\ee

By setting the variation of gravitino to zero we get the Killing
spinor equations, 
\be
\label{gravitino}
 \delta_\epsilon \psi_\mu =
   \nabla_\mu \epsilon - {1 \over 288} \l( \Gamma_{MNPQR} F^{NPQR}
  - 8 F_{MNPQ} \Gamma^{NPQ} \r) \epsilon = 0,
\ee
where
\be
 \nabla_\mu = \partial_\mu + {1 \over 4} \omega_\mu^{\hat{m}\hat{n}}
  \Gamma_{\hat{m}\hat{n}}
\ee
For each components they become
\be
\label{hm}
 &&\nabla_-\epsilon = 0 \\
\label{ha}
 &&\nabla_a\epsilon - {\mu \over 6} \Gamma_{\hat{a}} \Gamma_{\hm\hon\htw\hth}
   \epsilon = 0 \\
\label{hap}
 &&\nabla_{a'}\epsilon + {\mu \over 12} \Gamma_{\hat{a'}}
    \Gamma_{\hm\hon\htw\hth} \epsilon =0 \\
\label{hp}
 &&\nabla_+\epsilon + {\mu \over 12} \l( -\Gamma_{\hp\hm\hon\htw\hth}
   + 2 \Gamma_{\hon\htw\hth} \r) \epsilon = 0.
\ee
Spin connections can be calculated as
\be
 \omega_+{}^{\hat{a}\hm} &=& - \l( {\mu \over 3} \r)^2 y^a \\ \nn
 \omega_+{}^{\hat{a'}\hm} &=& - \l( {\mu \over 6} \r)^2 z^{a'}
\ee

From Eq. (\ref{hm}), we know that $\epsilon$ is independent of $-$.
From Eqs. (\ref{ha},\ref{hap}), we get
\be
\label{killing}
 \epsilon = \l(1 + {\mu \over 6} y^a \Gamma_{\hat{a}}\Gamma_{\hm\hon\htw\hth}
    - {\mu \over 12} z^{a'} \Gamma_{\hat{a'}} \Gamma_{\hm\hon\htw\hth}
    \r) \chi,
\ee
where $\chi=\chi(+)$.
Inserting this into Eq.(\ref{hp}), we get
\be
 \partial_+\chi + { \mu \over 12} \l( - \Gamma_{\hp\hm\hon\htw\hth}
  + 2 \Gamma_{\hon\htw\hth} \r) \chi =0.
\ee
The solution is
\be
 \chi=e^{{\mu \over 12} x^+ \Gamma_{\hp\hm\hon\htw\hth} } 
      e^{- {\mu \over 6} x^+ \Gamma_{\hon\htw\hth} } \epsilon_0,
\ee
where $\epsilon_0$ is a 32 components constant spinor.
\subsection{$AdS \times S$ in global coordinates}
Global metric for $AdS_7 \times S^4$ is
\be
 && ds^2 = R_{AdS}^2 \l( - \cosh^2 \rho d\tau^2 + d\rho^2 + \sinh^2 \rho
            d \Omega_5^2 \r) + R_S^2 dS_4^2 \nn \\
 && F_4 = 3 R_S^3 {\rm vol} (S^4),
\ee
where
\be
 dS_n^2 = d\theta_n^2 + \sin^2\theta_n^2 dS_{n-1}^2.
\ee
Here $R_{AdS} = 2 R_S$.
Penrose Limit is taken, with $R_S \rightarrow \infty$,
\be
 && \theta_i = {\pi \over 2} - {y_i \over R_S} (i=1,2,3) \nn \\
 && \rho = {z \over R_{AdS}} \nn\\
 && \tau = {\mu x^+ \over 6} + {6 x^- \over R_{AdS}^2 \mu} \nn\\
 && \theta = {\mu x^+ \over 3} - {3 x^- \over R_S^2 \mu} ,
\ee
where $\theta= \theta_4$.
We get
\be
\label{penrose}
 && ds^2 = -4 dx^+dx^- - \l( \l({\mu \over 3}\r)^2 y^2 + \l({\mu \over 6}\r)^2
       z^2 \r) dx^{+2} +
  d \vec{y}^2 + d\vec{z}^2 \nn\\
 && F_4 = \mu dx^+ \wedge dy^1 \wedge dy^2 \wedge dy^3.
\ee

Killing spinor equations  for global $AdS_7 \times S^4$ space are
\be
 &&\partial_\tau\epsilon + {1 \over 2} \sinh \rho 
   \Gamma_{\hat{\tau}\hat{\rho}} 
  \epsilon -{1 \over 2} \cosh \rho \Gamma_{\hat{\tau}} \Gamma_{\hat{*}} 
  \epsilon=0 \nn\\
 &&\partial_\rho \epsilon - {1 \over 2} \Gamma_{\hat{\rho}} \Gamma_{\hat{*}}
  \epsilon = 0 \nn \\
 &&\nabla_a \epsilon - {1 \over 2} \sinh \rho \,\, 
    e_a{}^{\hat{a}} \Gamma_{\hat{a}}
  \Gamma_{\hat{*}} \epsilon =0 \nn \\
 &&\nabla_{a'} \epsilon +{1 \over 2} e_{a'}{}^{\hat{a'}} \Gamma_{\hat{a'}}
  \Gamma_{\hat{*}} \epsilon =0,
\ee
where
\be
 \Gamma_{\hat{*}} = \Gamma_{\hat{\theta_1}\hat{\theta_2}\hat{\theta_3} 
   \hat{\theta_4}}.
\ee
Here $a= \phi_1, \cdots, \phi_5$ and $a'=\theta_1, \cdots, \theta_4$.
Vielbeins are defined such that 
$e_i^{}{\hat{i}} e_j{} ^{\hat{j}} \eta_{\hat{i}\hat{j}} = \bar{g}_{ij} $, 
where $ \bar{g}_{ij}$ is metric of unit sphere.
Solution is  
\be
\label{ads7s4killing}
 \epsilon= e^{{\rho \over 2} \Gamma_{\hat{\rho}} \Gamma_{\hat{*}} }
           e^{{\phi_1 \over 2} \Gamma_{\hat{\rho}\hat{\phi}_1} }
           \l(\prod_{k=1}^3 e^{ {\phi_{k+1} \over 2}
                 \Gamma_{\hat{\phi}_k\hat{\phi}_{k+1} }  } \r)
           e^{-{\theta_1 \over 2} \Gamma_{\hat{\theta}_1} \Gamma_{\hat{*}}}
           \l(\prod_{k=1}^3 e^{   {\theta_{k+1} \over 2}
                 \Gamma_{\hat{\theta}_k \hat{\theta}_{k+1}} } \r)
           e^{{\tau \over 2} \Gamma_{\hat{\tau}} \Gamma_{\hat{*}} }.
\ee

Global metric for $AdS_4 \times S^7$ is
\be
 && ds^2 = R_{AdS}^2 \l( - \cosh^2 \rho d\tau^2 + d\rho^2 + \sinh^2 \rho
            d \Omega_2^2 \r) + R_S^2 dS_7^2 \nn \\
 && F_4 = 3 R_{AdS}^3 \cosh \rho \sinh^2 \rho \sin \phi_1
           d\tau \wedge d\rho \wedge d\phi_1 \wedge d\phi_2,
\ee
where
\be
 dS_n^2 = d\theta_n^2 + \sin^2\theta_n^2 dS_{n-1}^2.
\ee
Here $R_{AdS} = 1/2  R_S$.
Penrose Limit is taken, with $R_S \rightarrow \infty$,
\be
 && \theta_i = {\pi \over 2} - {z_i \over R_S} (i=1,2,3) \nn \\
 && \rho = {y \over R_{AdS}} \nn\\
 && \tau = {\mu x^+ \over 3} + {3 x^- \over R_{AdS}^2 \mu} \nn\\
 && \theta = {\mu x^+ \over 6} - {6 x^- \over R_S^2 \mu} ,
\ee
where $\theta= \theta_7$.
We get the same metric as (\ref{penrose}). 

Killing spinor equations  for global $AdS_4 \times S^7$ space are
\be
 &&\partial_\tau \epsilon 
    + {1 \over 2} \sinh \rho \Gamma_{\hat{\tau} \hat{\rho}} \epsilon
    -{1 \over 2} \cosh \rho \Gamma_{\hat{\tau}} \Gamma_{\hat{*}} \epsilon=0
  \nn\\
 &&\partial_\rho \epsilon - {1 \over 2} \Gamma_{\hat{\rho}} \Gamma_{\hat{*}}
  \epsilon = 0 \nn \\
 &&\nabla_a \epsilon - {1 \over 2} \sinh \rho \,\, 
   e_a{}^{\hat{a}} \Gamma_{\hat{a}}
  \Gamma_{\hat{*}} \epsilon =0 \nn \\
 &&\nabla_{a'} \epsilon +{1 \over 2} e_{a'}{}^{\hat{a}'} \Gamma_{\hat{a}'}
  \Gamma_{\hat{*}} \epsilon =0,
\ee
where
\be
 \Gamma_{\hat{*}} = \Gamma_{\hat{\tau}\hat{\rho}\hat{\phi}_1 \hat{\phi}_2}.
\ee
Here $a= \phi_1, \cdots, \phi_2$ and $a'=\theta_1, \cdots, \theta_7$.
Vielbeins are defined such that 
$e_i^{}{\hat{i}} e_j{}^{\hat{j}} \eta_{\hat{i}\hat{j}} = \bar{g}_{ij} $, 
where $ \bar{g}_{ij}$ is metric of unit sphere.
Solution of Killing spinor equation is
\be
\label{ads4s7killing}
 \epsilon= e^{{\rho \over 2} \Gamma_{\hat{\rho}} \Gamma_{\hat{*}} }
           e^{{\phi_1 \over 2} \Gamma_{\hat{\rho}\hat{\phi}_1}}
           e^{{\phi_2 \over 2}
                 \Gamma_{\hat{\phi}_1 \hat{\phi}_2}}
           e^{-{\theta_1 \over 2} \Gamma_{\hat{\theta}_1} \Gamma_{\hat{*}}}
           \l(\prod_{k=1}^6 e^{{\theta_{k+1} \over 2}
                 \Gamma_{\hat{\theta}_k \hat{\theta}_{k+1} }} \r)
           e^{{\tau \over 2} \Gamma_{\hat{\tau}} \Gamma_{\hat{*}}}.
\ee
\end{appendix}

\end{document}